\documentclass{ws-mpla}
\usepackage{url}
\usepackage{color}

\begin{document}

\markboth{X. Qian and W. Wang}
{Reactor Neutrino Experiments: $\theta_{13}$ and Beyond}

%%%%%%%%%%%%%%%%%%%%% Publisher's Area please ignore %%%%%%%%%%%%%%
\catchline{}{}{}{}{}
%%%%%%%%%%%%%%%%%%%%%%%%%%%%%%%%%%%%%%%%%%%%%%%%%%%%%%%%%%%%%%%%%%%

\title{Reactor Neutrino Experiments: $\theta_{13}$ and Beyond}

\author{\footnotesize Xin Qian\footnote{email:xqian@bnl.gov}}

\address{Brookhaven National Laboratory, \\
Upton, NY, 11973, USA}

\author{\footnotesize Wei Wang\footnote{email:wswang@wm.edu}}
\address{Physics Department, College of William and Mary\\
Williamsburg, VA, 23187, USA}
\maketitle

\pub{Received (Day Month Year)}{Revised (Day Month Year)}

\begin{abstract}
We review the current-generation short-baseline reactor neutrino 
experiments that have firmly established the third neutrino 
mixing angle $\theta_{13}$ to be non-zero. 
The relative large value of $\theta_{13}$ 
(around 9$^\circ$) has opened many new and exciting opportunities
for future neutrino experiments. Daya Bay experiment with the
first measurement of $\Delta m^2_{ee}$ is aiming for a precision 
measurement of this atmospheric mass-squared splitting with a comparable 
precision as $\Delta m^2_{\mu\mu}$ from accelerator muon neutrino 
experiments. JUNO, a next-generation reactor neutrino experiment, 
is targeting to determine the neutrino mass hierarchy with 
medium baselines ($\sim$50 km). Beside these {\color{black} opportunities
  enabled by the large $\theta_{13}$}, the current-generation (Daya Bay,
Double Chooz, and RENO) and the next-generation (JUNO, RENO-50, 
and PROSPECT)  reactor experiments, with their unprecedented statistics, 
are also leading the precision era of the 3-flavor neutrino 
oscillation physics as well as constraining new physics beyond 
the neutrino Standard Model.

\end{abstract}

\ccode{PACS Nos.: include PACS Nos.}

\section{Introduction} % 3.0 pages adding abstract check

Reactor neutrinos have been playing a crucial role in the development 
of the Standard Model and the 3-flavor neutrino framework. In 1956, 
Cowan and Reines discovered neutrinos at the Savannah River reactor power 
plant in the U.S.~\cite{nu_dis}. In 2005, KamLAND experiment in Japan 
observed the neutrino oscillation in the solar 
sector~\cite{KamLAND_spec}. Their finding together with those from 
SNO experiment~\cite{SNO} in Canada firmly established the neutrino 
oscillation as the explanation of the solar neutrino 
puzzle.~\footnote{The solar
neutrino puzzle refers to a major discrepancy between 
measurements of the number of $\nu_e$ going through earth and 
that predicted by the standard solar model.} Most recently, 
Daya Bay experiment in China reported the discovery of non-zero
$\theta_{13}$, the third neutrino mixing angle, with a significance 
$>$5$\sigma$ in 2012~\cite{dayabay}. The non-zero $\theta_{13}$ opens
the gateway to access two (out of three) remaining unknown parameters
in the neutrino Standard Model: the neutrino mass hierarchy and the leptonic 
CP phase $\delta_{CP}$.~\footnote{The other unknown parameter is the mass of the
lightest neutrino.}

\begin{figure}[htp]
\begin{centering}
\includegraphics[width=0.59\textwidth]{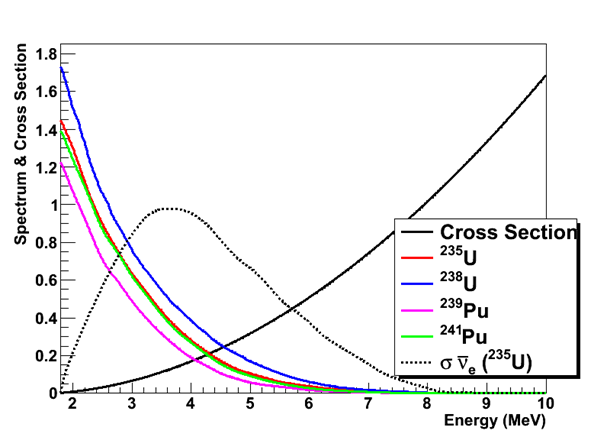}
\includegraphics[width=0.39\textwidth]{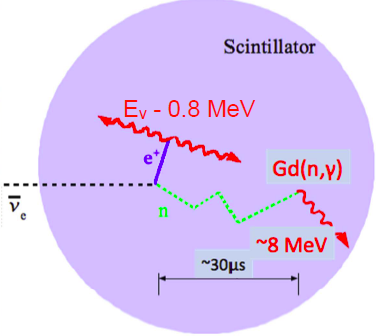}
\par\end{centering}
\caption{\label{fig:IBD} (Left) $\bar{\nu}_e$ energy spectra 
(four curves with negative slopes) for $^{235}$U, 
$^{238}$U, $^{239}$Pu, and $^{241}$Pu are shown. The curve with positive
slope represents the cross section of the inverse beta decay (IBD)
process. The convoluted IBD spectrum, seen in experiments,  
is shown as the dotted line. (Right) The detecting principle of 
IBD is shown. }
\end{figure}

Reactor is essentially a pure electron antineutrino $\bar{\nu}_e$ source 
with an average of six $\bar{\nu}_e$ produced per fission along the 
$\beta$-decay chain of fission products.~\footnote{
There is a small component of the electron neutrino $\nu_e$ with energy $\sim$0.1 
MeV from the neutron activation of shielding materials.}
For a 1 GW reactor thermal power, about 2$\times$10$^{20}$ $\bar{\nu}_e$
are emitted every second isotropically. 
%with an average energy $\sim$200 MeV
%released per fission. 
Inside the reactor core, the fission process
is maintained by neutrons produced through the fission 
of $^{235}$U nucleus. The condition is adjusted so that only one 
neutron out of the few generated by the $^{235}$U fission can induce 
a new fission. Meanwhile, a portion of the neutrons are captured by the 
$^{238}$U producing new fissile isotopes: $^{239}$Pu and $^{241}$Pu.
These four isotopes are main sources of $\bar{\nu}_{e}$. The $\bar{\nu}_e$
energy spectra are shown in the left panel of Fig.~\ref{fig:IBD}. 

As shown in the right panel of Fig.~\ref{fig:IBD}, reactor $\bar{\nu}_e$ is 
detected through the inverse beta decay (IBD) reaction with free 
protons: $\bar{\nu}_e + p \rightarrow e^{+} + n$. An IBD event is 
a pair of coincident signals consisting
%a coincidence 
%signal consisting 
i) a prompt signal induced by the positron ionization and 
annihilation inside the detector (such as a liquid scintillator LS detector) 
and ii) a delay signal produced 
by the neutron capture on proton or nucleus (such as Gd). In particular, the 
neutron capture on Gd would release multiple gammas with a total energy 
$\sim$8 MeV. With 0.1\% Gd doped LS, the mean time between the prompt 
and the delay signal is about 30 $\mu$s. Due to the time-correlation
nature, IBD can be easily distinguished from radioactive backgrounds which mostly consist of only a single signal. 
Furthermore, the energy of the prompt signal is directly linked 
to the neutrino energy: $E_{\nu} \approx E_{prompt} + 0.78~$MeV. This is in 
particular an attractive feature for measurements of neutrino oscillations that 
require knowledge of the neutrino energy. The left panel of Fig.~\ref{fig:IBD}
shows the total cross section of the IBD process and the convoluted energy spectrum 
in reactor experiments. 

The current-generation reactor neutrino experiments including Daya Bay, 
Double Chooz, and RENO are designed to measure the third neutrino
mixing angle $\theta_{13}$ in the neutrino mixing (commonly referred to 
as the Pontecorvo-Maki-Nakagawa-Sakata or PMNS in short~\cite{ponte1,ponte2,Maki}) 
matrix. 
The survival probability of $\bar{\nu}_{e}$ with energy $E_{\nu}$ at a distance
$L$ is written as:
%The neutrino oscillation probability for the $\bar{\nu}_{e}$ disappearance
%is written as:
\begin{equation}\label{eq:osc}
P_{\bar{\nu}_e\rightarrow \bar{\nu}_e} = 1 - \sin^22\theta_{13} \cdot 
\sin^2\left( \Delta m^2_{ee} \cdot \frac{L}{E_{\nu}}  \right) - 
\cos^4\theta_{13} \cdot \sin^22\theta_{12} \cdot \sin^2\left( \Delta m^2_{21} \cdot \frac{L}{E_{\nu}}\right).
\end{equation}
Here, $\Delta m^2_{ij} := m^2_{i}-m^2_{j}$ are neutrino 
mass-squared differences. From Ref.~\cite{PDG}, we have 
$\Delta m^2_{21} \approx 7.6\times 10^{-5}~$eV$^{2}$ and 
$\Delta m^2_{ee} \approx 2.4\times 10^{-3}~$eV$^{2}$ that is
a combination of $\Delta m^2_{32}$ and $\Delta m^2_{31}$~\cite{Nuno}. 
$\theta_{12}\sim
32^{\circ}$~\cite{PDG} is the second neutrino mixing angle. 
%$L$ is the distance neutrino travels. 
From Eq.~\eqref{eq:osc}, it is easily seen that the 
$\bar{\nu}_{e}$ disappearance is a very clean channel to access $\theta_{13}$.
Unlike the $\nu_{\mu}\rightarrow\nu_e$ appearance channel, the disappearance
channel is not sensitive to the mass hierarchy (sign of $\Delta m^2_{32}$) 
through the matter effect and is immune to the unknown CP phase 
$\delta_{CP}$ in the PMNS matrix. 

\begin{figure}[htp]
\begin{centering}
\includegraphics[width=0.7\textwidth]{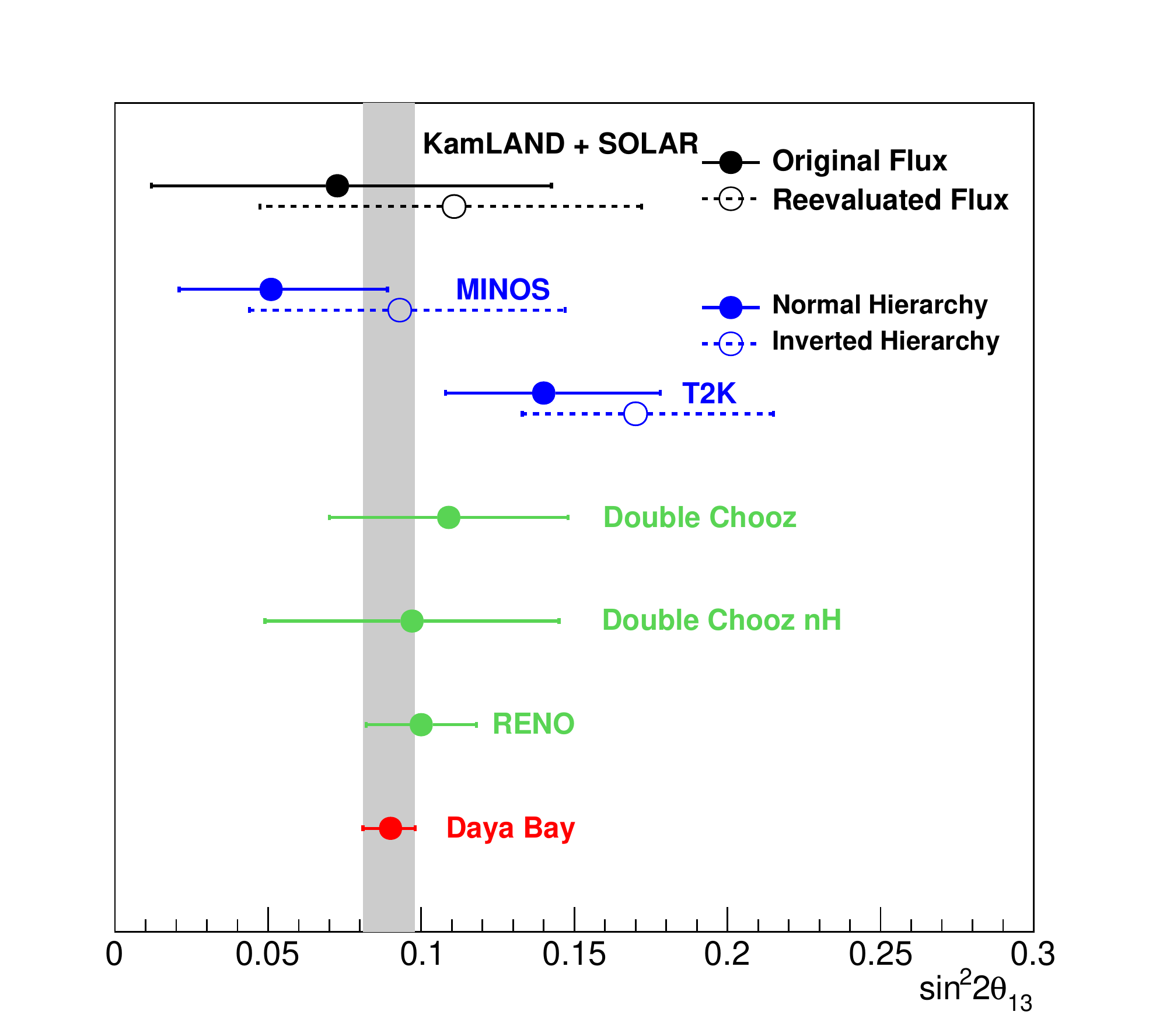}
\par\end{centering}
\caption{\label{fig:global} Global status of $\sin^22\theta_{13}$ including 
KamLAND+SOLAR~\protect\cite{Fogli:2011qn}, MINOS~\protect\cite{Evans:2013pka}, 
T2K~\protect\cite{Abe:2013hdq}, Double Chooz~\protect\cite{Abe:2013sxa}, 
RENO~\protect\cite{2013arXiv1312.4111S}, and Daya Bay~\protect\cite{An:2013zwz}.}
\end{figure}

The first attempt to measure $\theta_{13}$ is by
CHOOZ~\cite{Chooz1,Chooz2} and Palo Verde~\cite{PaloVerde} experiments in
late 1990s and early 2000s. No oscillations were observed and 
an upper limit of $\sin^22\theta_{13}<0.12$ was set at 90\% C.L. by Chooz.
In 2011, there were several hints suggesting a non-zero $\theta_{13}$. 
The first one is from the
tension~\cite{Fogli:2011qn} between the KamLAND $\bar{\nu}_{e}$ disappearance
measurement and the solar measurements (e.g. ratio of $\nu_e$ to the neutral current
interactions from SNO). Subsequently, MINOS~\cite{minos} and T2K~\cite{t2k} reported their
searches of $\nu_{\mu}$ to $\nu_e$ oscillation that is also sensitive to
$\theta_{13}$. In particular, T2K~\cite{t2k} disfavored the $\theta_{13}=0$
hypothesis at 2.5$\sigma$. In early 2012, Double Chooz~\cite{dc} reported
that the $\theta_{13}=0$ hypothesis was disfavored at 1.6$\sigma$ with only
the far detector. 
A $>$5$\sigma$ discovery of non-zero $\theta_{13}$ was finally made by 
Daya Bay in March 2012~\cite{dayabay}. 
One month later, RENO confirmed the Daya Bay discovery with a 
$4.9\sigma$ significance~\cite{RENO}. Non-zero $\theta_{13}$ was firmly 
established. Fig.~\ref{fig:global} shows the current-global status of 
$\sin^22\theta_{13}$ measurements compiled with the latest results 
from each experiment. 

In the following, we review current-generation reactor experiments 
and present an outlook of future reactor experiments. As shown in 
Eq.~\eqref{eq:osc}, a non-zero $\theta_{13}$ will lead to $\bar{\nu}_e$
disappearance at $\sim$2 km corresponding to the oscillation length of 
the atmospheric mass-squared difference at $E_{\nu}=$4 MeV (the peak of the reactor IBD energy 
spectrum). In practice, the search for such a deficit with a single detector 
is limited by the theoretical uncertainty of the antineutrino flux, which 
was considered to be larger than the speculated deficit when the current-generation 
experiments were designed. In order to suppress this 
uncertainty, the current generation experiments adopt the ratio strategy~\cite{ratio},
in which identical detectors were deployed close to (near detectors at 0.3-0.5 km)
and further away from (far detectors at 1-2 km) reactor cores. This dual-detector configuration is essential to achieve high precision measurements of 
$\sin^22\theta_{13}$.

The large size of $\theta_{13}$ has generated new opportunities which 
include the resolution of the neutrino mass hierarchy at 
medium-baseline reactor oscillation (MBRO) experiments. 
We will provide a brief review of MBRO principle and the JUNO experiment. 
Furthermore,  a new evaluation of the reactor antineutrino flux revealed a discrepancy
of about 5.7\% between the calculation and very short baseline ($<100$ m) measurements~\cite{anom}.
This deficit is usually referred to as the "reactor anomaly".
 An updated analysis, including kilometer-scale reactor experiments and improved treatment 
of correlations among experiments suggested a smaller discrepancy of 4.1\%~\cite{anom1}. 
Recently, authors of Ref.~\cite{anom2} suggested that the uncertainty of 
reactor neutrino flux should be larger than 5\%.
To provide a definite answer, a new generation of very short-baseline (VSBL) 
reactor neutrino experiments have been proposed to address the "reactor anomaly". 
We will briefly review one U.S. effort, PROSPECT.

\section{Daya Bay Reactor Neutrino Experiment} % 5 pages 
\subsection{Design of the experiment}
\begin{figure}[htp]
\centering
\includegraphics[width=0.49\textwidth]{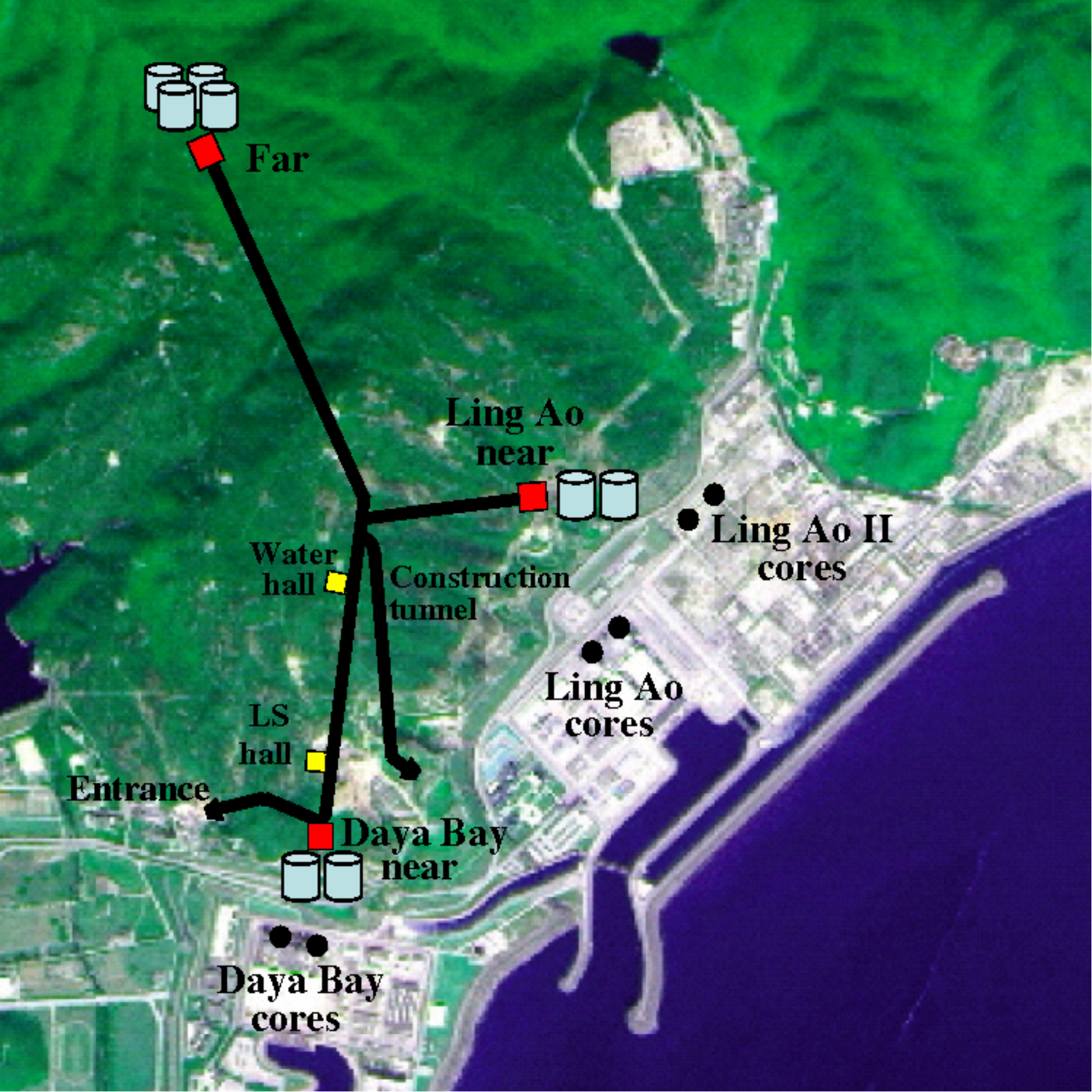}
\includegraphics[width=0.49\textwidth]{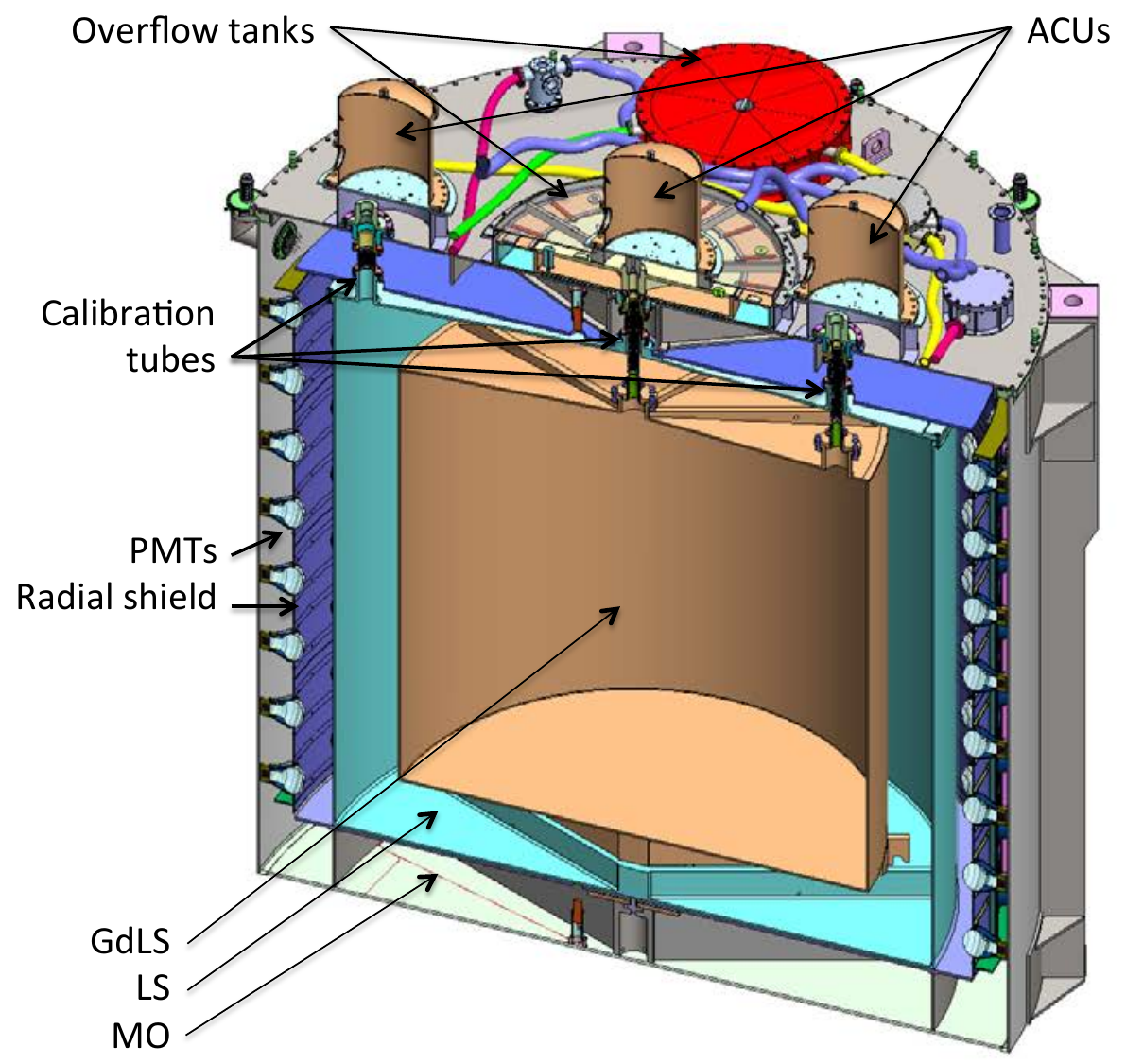}
\caption{\label{fig:dyb_map} Left panel shows the layout and the map of the
  Daya Bay experiment and the hosting Daya Bay plant campus. Right panel
  shows the structure of the Daya Bay antineutrino detector~(AD). The Daya
  Bay ADs are equipped with three automated calibration units~(ACUs), two for
  the Gd-LS volume and one for the LS volume. }
\end{figure}

Daya Bay Reactor Neutrino Experiment is located on the campus of the
Daya Bay nuclear reactor power plant in South China. As shown in the 
left panel of Fig.~\ref{fig:dyb_map}, the plant hosts six
reactor cores whose locations are grouped into two clusters: the Daya Bay cluster
that includes Daya Bay I and II cores and the Lingao cluster that
includes Lingao I through IV cores. The total thermal power is about 17.4 GW.
To monitor the antineutrino fluxes from these two
clusters, Daya Bay has designed two near underground sites, the
Daya Bay site (363 m from Daya Bay cores) and the Lingao site ($\sim$500 m 
from Lingao cores). Each near underground site hosts two
antineutrino detectors (ADs). The far site is located at a position that 
maximizes the sensitivity to $\theta_{13}$, considering the overburden and
geological conditions for the construction of an underground lab. 
The average baseline is about $\sim 1.7$ km. The near-far arrangement 
of the experiment guarantees that the reactor antineutrino flux uncertainty 
is largely canceled. The far site hosts four ADs which pair with 
the four ADs of the two near sites, providing a maximal cancellation
of detector effects. The effective vertical overburdens are 250, 265, and 860 
water-equivalent meters for EH1, EH2, and EH3, respectively.
  
%design of the AD
Right panel of Fig.~\ref{fig:dyb_map} shows the schematic view of the 
antineutrino detector~(AD)~\cite{Band:2013osa}. Daya Bay adopts a three-zone cylindrical 
shaped design, with inner, middle, and outer layer containing 20 t 
Gd-doped (0.1\% in weight) LAB-based liquid scintillator (GdLS), 22 t liquid scintillator 
(LS), and 40 t mineral oil, 
respectively. 192 8-inch PMTs are installed on each AD. The photo-cathode
coverage is about 8\%, which is further enhanced by the top and bottom
optical reflectors to about 12\%. Three automated calibration units (ACUs)~\cite{Liu:2013ava}
are equipped. Each ACU contains four sources: i) a LED for the PMT 
gain/timing calibration, ii) a $^{68}$Ge source for the IBD threshold 
calibration, iii) a $^{60}$Co source for the determination of the overall 
energy scale, and iv) a $^{241}$Am-$^{13}$C neutron source to understand
neutron captures on Gd and to determine the H to Gd neutron capture ratio 
in the target (GdLS) region. 

ADs are placed inside high purity water to reduce radioactive backgrounds
from the environment. Each water pool is divided into two optically 
separated regions: the inner water pool (IWS) and the outer water
pool (OWS). With PMTs installed, each region of water pool also 
operates as an independent water Cerenkov detector. The detection 
efficiencies for cosmic muons are measured to be 99.7\% and 97\% 
for the IWS and OWS~\cite{DayaBay:2012aa}, respectively. A layer of 
resistive plate chamber (RPC) is further installed above each 
water pool as an additional muon tagging detector. 

\subsection{Signal and Backgrounds}
The IBD events in Daya Bay are selected with the following cuts~\cite{dayabay,An:2013zwz}: i)
the energy of the prompt signal is between 0.7 and 12 MeV, 
ii) the energy of the delay signal is between 6 and 12 MeV,
and iii) the time difference between the prompt and the delay signal
is between 1 and 200 $\mu$s. In addition, a multiplicity 
cut is applied to remove energy ambiguities in the prompt 
signal. The overall selection efficiency is about 80\%. 
In order to suppress cosmogenic backgrounds, three 
types of muon vetos are applied to the delay signal: i) the 
water pool muon: from 2 $\mu$s before to 600 $\mu$s after
the water pool signal, ii) the AD shower muon ($>3\times 10^{5}$
photoelectrons): from 2 $\mu$s before to 0.4 s after
the AD shower, and iii) the AD non-shower muon ($>$20 MeV):
from 2 $\mu$s before to 1.4 ms after the AD signal. 

There are in total five backgrounds~\cite{dayabay,An:2013zwz}. 
The first one is the accidental background, which consists two 
uncorrelated single signals, and can be calculated with negligible systematic
uncertainties with the measured rate of single signal.
It is about 1.7\% and 4.6\% of IBDs at near and 
far sites, respectively.  
The second one is the correlated background induced by
the Am-C neutron source inside ACU. The energetic neutron could 
go through an inelastic scattering with an Fe nuclei emitting 
a gamma and then followed by an Fe capture emitting another gamma. 
The correlated background occurs when both gammas enter the
AD. The rate of this background was estimated by the simulation 
and further validated by a special run with a strong Am-C
source. It is about 0.03\% and 0.3\% of IBDs at
near and far site, respectively. The relative uncertainty is about 30\%.
The third background is $^{9}$Li and $^{8}$He generated by 
cosmic muons. They are both long-lived isotopes which can 
not be excluded by muon vetos. They would firstly 
go through beta-decay process (prompt). The daughter nucleus could emit a
neutron (delay). The rates can be directly measured by tagging
muons. They are about 0.35\% and 0.2\% of IBDs at near and 
far site, respectively. The uncertainties are about 30-50\%. 
The fourth background is the fast neutrons produced by cosmic 
muons. The fast neutrons could go through an elastic scattering 
with proton (prompt) and followed by a capture (delay). They
can also be directly measured by tagging the muon. It is 
about 0.13\% and 0.1\% for the near and far sites, respectively. 
The uncertainty is about 30\%. The last background ($\alpha$-N)
is induced by internal radioactive backgrounds and is below
0.1\%. Besides backgrounds, the detector related uncertainties entering
into the oscillation analysis are dominated by the 
0.12\% from the 6 MeV delay energy cut  and $\sim$0.1\% from 
the H to Gd neutron capture ratio. The reactor
related uncertainties, suppressed by near/far ratios, are
$\sim$0.04\%. 
 
\subsection{Detector Energy Calibration}
% describe the calibration system and the detector energy model
% include the energy model plot and

Reactor IBD spectrum covers the antineutrino energy range from 
from 1.8 MeV to $\sim$8 MeV. The analysis of the spectral distortion between 
the near and far detectors can provide additional information 
on $\sin^22\theta_{13}$ as well as new information on 
$\Delta m^2_{ee}$. In this analysis, understanding the
absolute energy response of the prompt positron signal is crucial. 
The LS energy response in Daya Bay is illustrated in the following. 
First, a positron with a kinetic energy $E_{true}$ would deposit
$E_{dep}$ into the LS through the ionization and the annihilation processes. 
Second, some of the deposited energy will convert to scintillation light 
and Cerenkov radiation. Due to the quenching process of the LS, the conversion between 
$E_{dep}$ and scintillation light is not linear. In
addition, Cerenkov radiation emerges only when the particle
is above the Cerenkov threshold of the LS. Total light collected by 
PMTs including both scintillation and Cerenkov lights is referred to as 
the visible energy $E_{vis}$. Finally, the readout electronics
will convert $E_{vis}$ into the reconstructed energy $E_{rec}$
used in the oscillation analysis. The conversion between $E_{true}$
and $E_{vis}$ is referred to as the scintillator nonlinearity. The 
conversion between $E_{vis}$ and $E_{rec}$ is referred to as the
electronics nonlinearity. 

In Daya Bay, the scintillator energy model is based on the LS response to
electron. The response to gamma is connected to that to electron through
a GEANT4 simulation.~\footnote{Gammas deposit energy in LS via
  electrons/positrons produced through Compton scatterings and pair
  productions.} The detector response to the 
ionization energy loss of positron is assumed to
be the same as that to electron. There are two additional 0.511 MeV gammas
from the positron annihilation. Two approaches are used to parametrize the LS
response to electron: i) Birks law for scintillation plus Cerenkov
contributions and ii) direct parametrization inspired by i). The functional
form of the electronics nonlinearity is inspired by the Monte Carlo simulation
of the electronics. 

\begin{figure}[htp]
\centering
\includegraphics[width=0.7\textwidth]{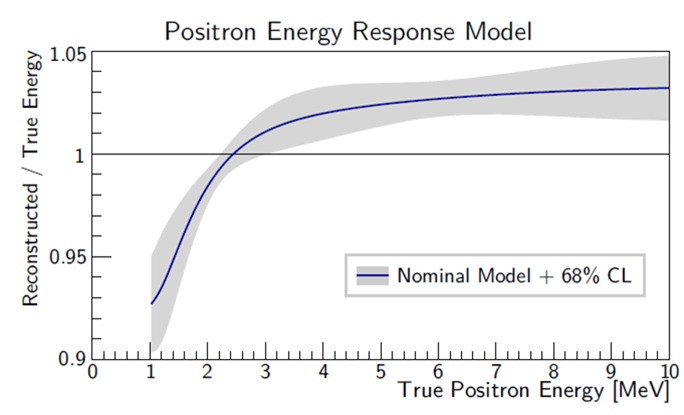}
\caption{\label{fig:energy}The Daya Bay positron energy model taken from 
Ref.~\protect\cite{An:2013zwz}.}
\end{figure}

The energy model is constrained by the calibration with gamma sources
and the well-known $^{12}$B beta decay continuous spectrum. The gamma 
sources include i) regularly deployed radioactive calibration sources:
$^{68}$Ge, $^{60}$Co, and $^{241}$Am-$^{13}$C,  ii) additional 
radioactive sources deployed during a special calibration period:
$^{137}$Cs, $^{54}$Mn, $^{40}$K, $^{241}$Am-$^{9}$Be, and Pu-$^{13}$C,
and iii) singles during regular physics data taking: $^{40}$K, 
$^{208}$Tl, and n-H capture. The $^{12}$B that are produced by the muon spallation 
inside the scintillator are selected by tagging the muon signal. 
In addition, the model is further checked with $\alpha$ peaks from $^{210}$Po,
$^{219}$Rn, $^{212}$Po, $^{214}$Po, and $^{215}$Po
and continuous beta-decay spectra from $^{212}$Bi, $^{214}$Bi, and $^{208}$Tl.
Several models are independently developed by different analysis teams, 
and the final positron energy model is conservatively taken as linear 
combinations of five energy models. Fig.~\ref{fig:energy} shows the 
Daya Bay positron energy model and its uncertainty band~\cite{An:2013zwz}.

\subsection{$\sin^22\theta_{13}$ and $\Delta m^2_{ee}$ }

\begin{figure}[htp]
\centering
\includegraphics[width=\textwidth]{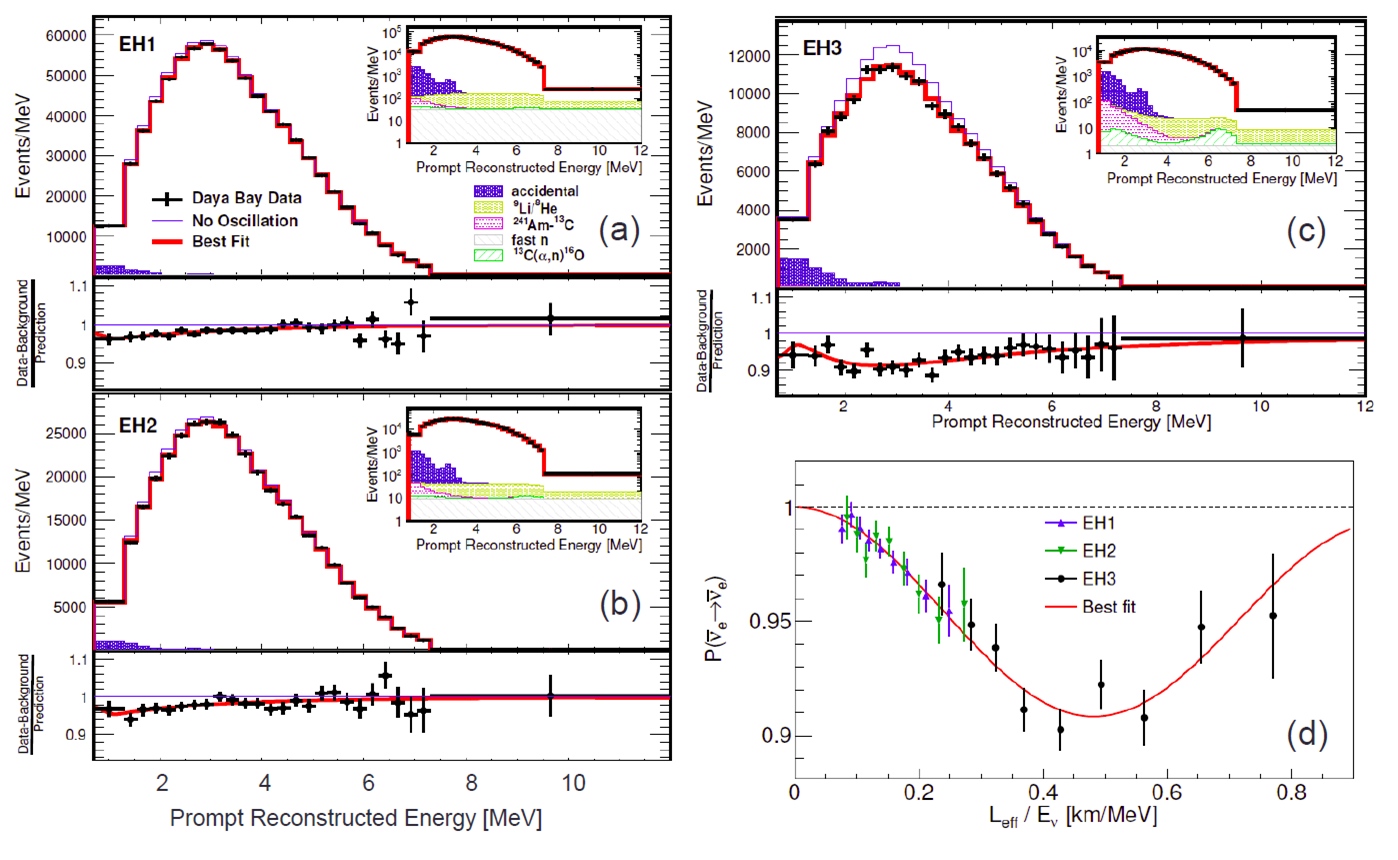}
\caption{\label{fig:dyb_result} 
6-AD data 
and best-fit spectra in three experimental halls are shown as a), b),
and c). The electron antineutrino 
survival probability vs. the effective propagation distance $L_{eff}$ over the antineutrino
energy $E_{\nu}$ is shown in d). An effective detector-reactor distance
$L_{eff}$ is calculated for each experimental hall by treating the multi-core
oscillated flux as it is from a single reactor core. Plots are taken from Ref.~\protect\cite{An:2013zwz}. }
\end{figure}

The oscillation analysis is based on the standard $\chi^2$ method
with Poisson statistics. The variations in systematics are included
as penalty terms in the $\chi^2$. The rate-only analysis~\cite{dayabay} 
of full 6-AD data ($\sim$215 days of data taking) yields $\sin^22\theta_{13}
=0.089\pm0.009$ with $\chi^2/$NDF=0.48/4. In this analysis, the constraint
of $\Delta m^2_{ee}$ is added according to the MINOS measured 
$\Delta m^2_{\mu\mu}=2.41^{+0.09}_{-0.10}\times
10^{-3}~$eV$^2$~\cite{minos_update}.
In the Daya Bay rate+shape analysis, uncertainties in the reactor flux
predictions are based on 
Ref.~\cite{VonFeilitzsch:1982jw,Schreckenbach:1985ep,Hahn:1989zr,Huber:2011wv,Vogel:1980bk,Mueller:2011nm}. 
The constraint is implemented as a covariance 
matrix in the penalty terms. The rate+shape analysis yields 
$\sin^22\theta_{13}=0.090^{+0.008}_{-0.009}$ and 
$\Delta m^2_{ee}=2.59^{+0.19}_{-0.20}\times10^{-3}~$eV$^2$ with 
$\chi^2/$NDF=162.7/153~\cite{An:2013zwz}. The $\Delta m^2_{ee}$ result
corresponds to $\Delta m^2_{32}=2.54^{+0.19}_{-0.20}\times 10^{-3}~$eV$^2$
assuming the normal mass hierarchy or $\Delta
m^2_{32}=-2.64^{+0.20}_{-0.19}\times 10^{-3}~$eV$^{2}$ assuming the inverted
mass hierarchy. These results are consistent with the $\Delta m^2_{\mu\mu}$
measured in MINOS ($\Delta m^2_{32}=2.37^{+0.09}_{-0.09}\times 10^{-3}~$eV$^2$
assuming the normal mass hierarchy or $\Delta
m^2_{32}=-2.41^{+0.11}_{-0.09}\times 10^{-3}~$eV$^{2}$ assuming the inverted
mass hierarchy)~\cite{minos_update}.
Fig.~\ref{fig:dyb_result} shows the best-fit IBD spectra in all three 
experimental halls. In addition, we show the electron antineutrino 
survival probability vs. the effective propagation distance $L_{eff}$ over 
$E_{\nu}$.

\subsection{Outlook}

\begin{figure}[htp]
\begin{centering}
\includegraphics[width=0.49\textwidth]{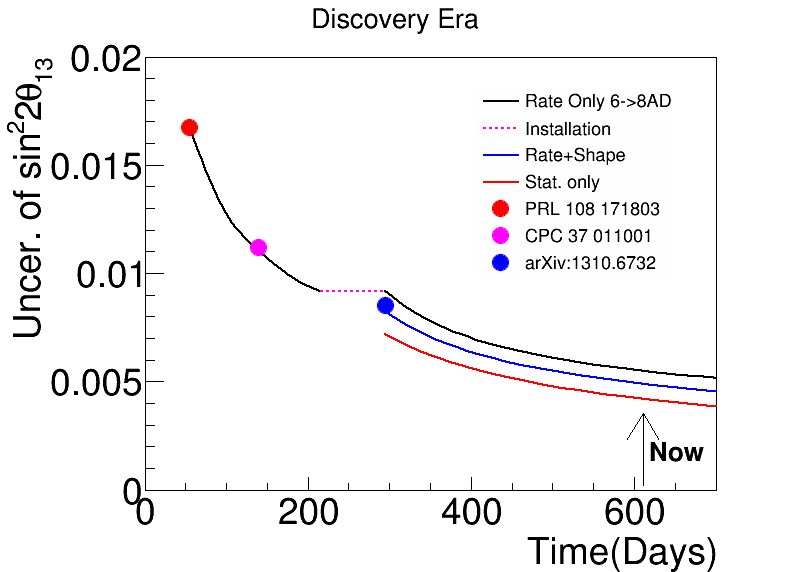}	
\includegraphics[width=0.49\textwidth]{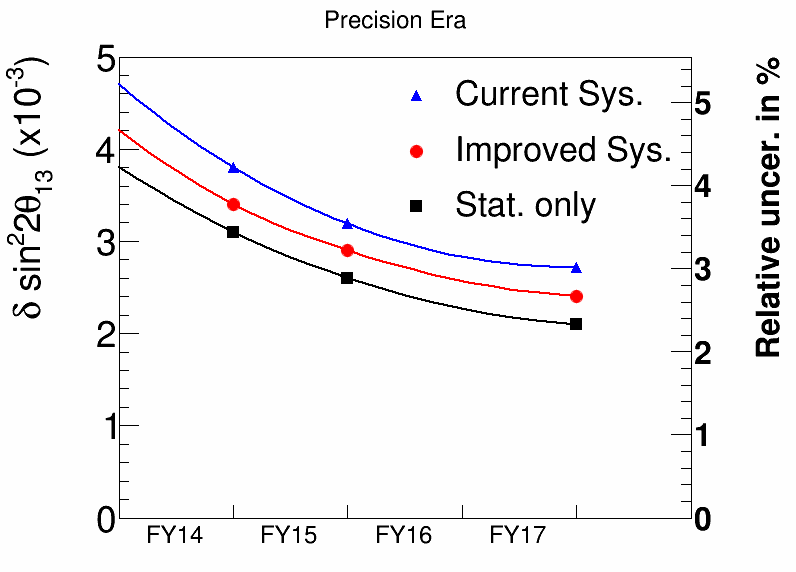}
\par\end{centering}
\caption{\label{fig:dyb_sen1}  
(Left) Uncertainty on the 
Daya Bay measurement of $\sin^22\theta_{13}$ over time under
different assumptions. (Right) Uncertainty on 
the Daya Bay measurement of $\sin^22\theta_{13}$ in the precision
era (FY14-FY17) under different assumptions. Plots are taken from Ref.~\protect\cite{dyb_run}.}
\end{figure}

\begin{figure}[htp]
\begin{centering}
\includegraphics[width=0.49\textwidth]{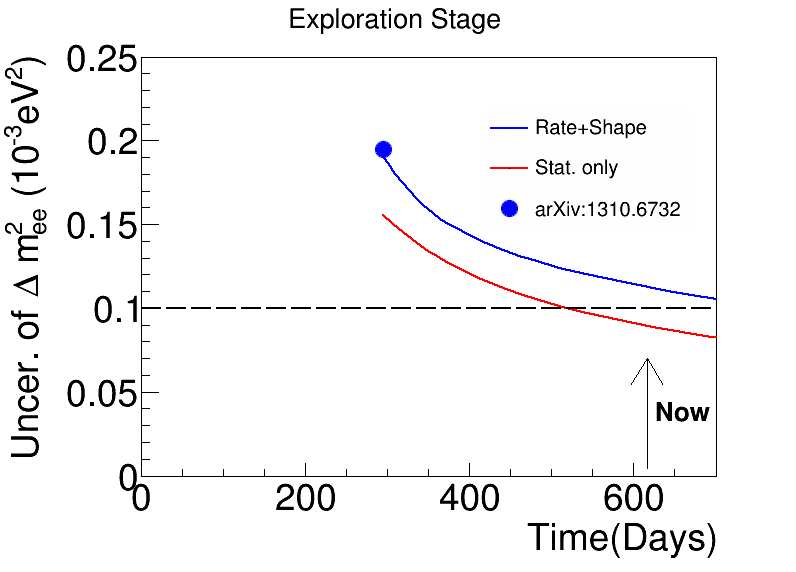}	
\includegraphics[width=0.49\textwidth]{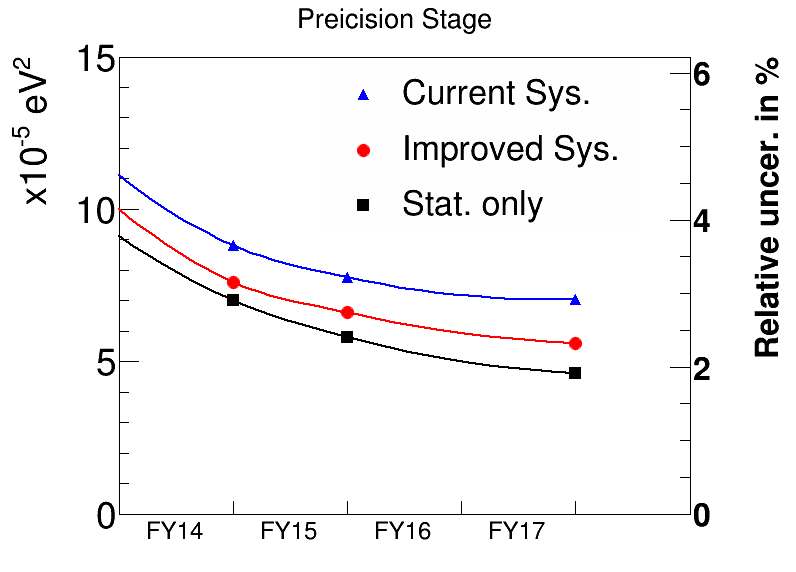}
\par\end{centering}
\caption{\label{fig:dyb_sen2} 
(Left) The expected Daya Bay uncertainty on $\Delta m^2_{ee}$ with 
existing systematic uncertainties and the statistical-only uncertainty. The
horizontal dash line is the current MINOS uncertainty in $\Delta
m^2_{\mu\mu}$. (Right) The evolution of the Daya Bay's uncertainty on
$\Delta m^2_{ee}$ is shown for a few different scenarios in the precision
era. Plots are taken from Ref.~\protect\cite{dyb_run}. }
\end{figure}

Daya Bay is entering the precision phase with data taking through 
2017~\cite{dyb_run}. As shown in Fig.~\ref{fig:dyb_sen1},
the $\sin^22\theta_{13}$ will be measured to better than 3\% 
(an absolute uncertainty of 0.003). It will stand as the world's most
precise measurement for the  foreseeable future. The precision measurement
of $\sin^22\theta_{13}$ will also improve the measurement of other mixing
parameters  by accelerator experiments. Furthermore, the comparison of the
precision measurement of $\sin^22\theta_{13}$ in reactor experiments and
that from accelerator experiments (such as LBNE~\cite{Adams:2013qkq}) will 
be one of the most stringent unitarity tests of the PMNS neutrino
mixing matrix~\cite{Qian:2013ora}. This is a crucial test of 
the standard 3-flavor neutrino framework in analogy to the 
unitarity test of the quark-mixing (CKM) matrix. 

As shown in Fig.~\ref{fig:dyb_sen2}, Daya Bay will reach a precision 
of $\Delta m^2_{ee}$ to about 2.5\%, which will be competitive
with that of $\Delta m^2_{\mu\mu}$ currently set by MINOS. This will be another 
stringent test to the 3-flavor neutrino framework. 
In addition, the comparison of $\Delta m^2_{ee}$ and
$\Delta m^2_{\mu\mu}$ will provide additional information
regarding to the neutrino mass hierarchy. 

Daya Bay will have the largest sample of reactor 
IBD events with more than one million interactions. Such a large 
sample of IBDs will provide excellent opportunities to study the 
reactor antineutrino spectrum as well as a precision flux
measurement at a distance of $\sim$360 m. In addition,
with the unique three sites configuration (e.g. three baselines), 
Daya Bay allows a competitive search for a sterile neutrino in the 
mass-squared splitting range of 0.001-0.2 eV$^2$ with 
excellent sensitivities.

\section{RENO} % 2 pages 
\begin{figure}[htp]
\centering
\includegraphics[height=1.85in]{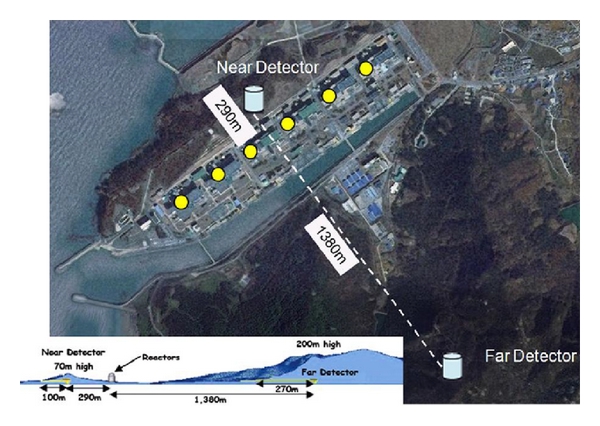}
\includegraphics[height=2.1in]{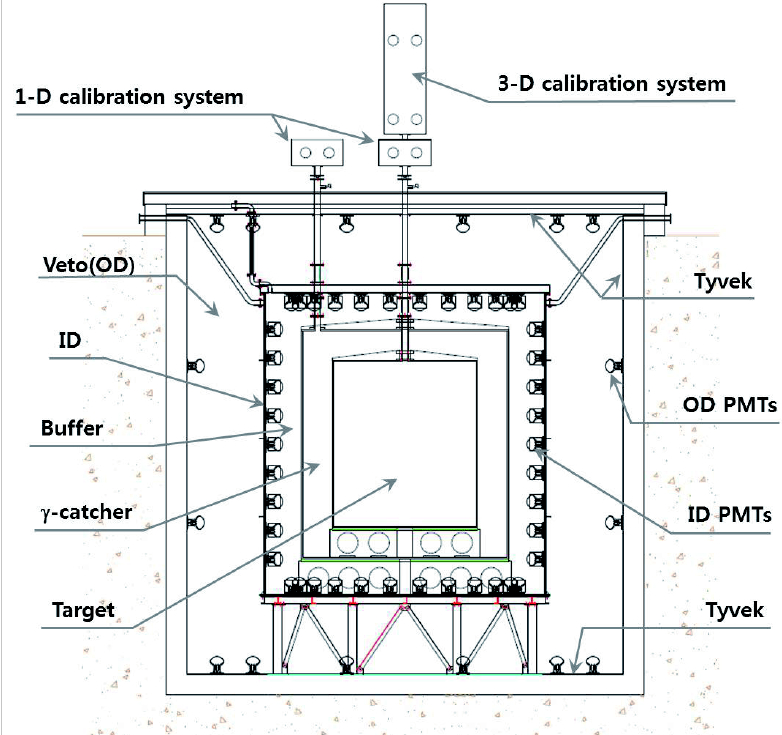}
\caption{\label{fig:reno-map-det}The left panel shows the Yonggwang nuclear
  reactor complex. The six reactors are equally spaced on one line. RENO
  near site is located $\sim$290 m away from the reactor complex center and
  the far site is $\sim$1,380 m away. The overburden is 70 m rock (185 meter water 
  equivalent or mwe) at the near
  site and 200 m (530 mwe) at the far site. The right panel shows RENO's detector
  system. A 3-D calibration system is installed in the GDLS. 
  The gamma catch region is equipped with a 1-D calibration system
  which moves calibration sources along the vertical direction. }
\end{figure}

%Besides Daya Bay, there are two competing short-baseline reactor experiments
%RENO and Double Chooz~\cite{}. 
RENO is another current-generation short-baseline reactor neutrino
experiment aiming at measuring the value of $\sin^2\theta_{13}$ and it has
confirmed the Daya Bay discovery of non-zero $\theta_{13}$ with a near
5-$\sigma$ confidence level. The
experiment is built near the Yonggwang nuclear power plant in South Korea. The
total thermal power of the six reactor cores is about 16.4
GW. The baseline distribution of RENO is shown in
Fig.~\ref{fig:reno-map-det}. With a symmetric core configuration, 
RENO has one near site and one far site to suppress the reactor 
antineutrino flux uncertainty. The distance between RENO's near
site and the geometrical center of reactor cores is $\sim$290 m.
For the far site, the distance is $\sim$1,380 m.
The arrangement of the RENO detector system has taken a
similar approach as the Daya Bay one: a three-zone LS antineutrino 
detector is nested in a muon veto system. The RENO LS is also LAB-based. 
The target zone contains 16.1 t 0.1\% Gd-doped LS. 
%RENO is the first short-baseline reactor experiment that has confirmed the
%non-zero $\theta_{13}$ value measured by Daya Bay~\cite{}.
%value of $\sin^2 2\theta_{13}$ 
%~(Reactor Experiment of Neutrino Oscillation)

RENO had collected $\sim$800 live days of data by the end of 2013 and its
statistical uncertainty has surpassed the systematic one. The latest result
based on the rate analysis of RENO is 
$\sin^22\theta_{13}=0.100\pm0.010~(\text{stat.})\pm0.015~(\text{sys.})$~\cite{2013arXiv1312.4111S}.

\section{Double Chooz} % 2 pages 
\begin{figure}[htp]
\centering
\includegraphics[height=1.8in]{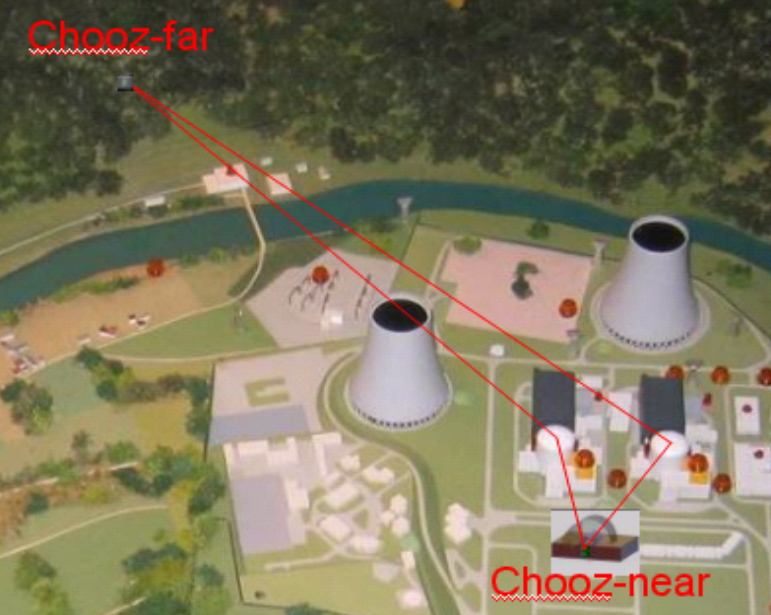}
\includegraphics[height=2.1in]{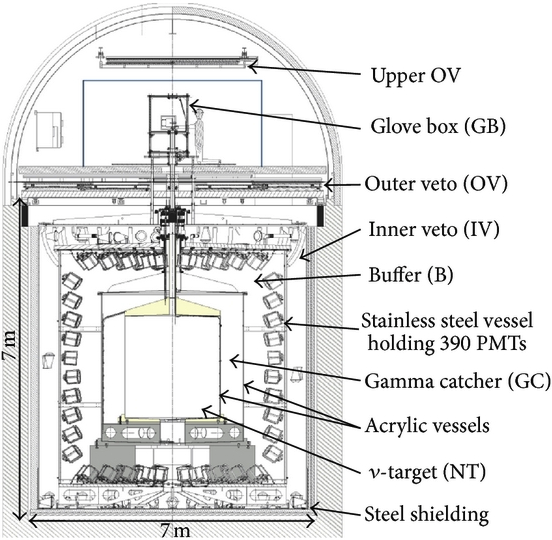}
\caption{\label{fig:dc-map-det} (Left) The 
  two nuclear reactors and the two detector locations of Double Chooz are shown. 
Double Chooz's near site is located $\sim$410 m away and
  the far site is the original Chooz site which is $\sim$1,067 m away. The
  overburden is 115 mwe and 300 mwe at the near and far site, respectively. (Right) The Double Chooz detector system.}
\end{figure}

Double Chooz experiment is built upon the previous generation Chooz
experiment that set the best $\sin^22\theta_{13}$ upper limit previously.
The Double Chooz design expands the Chooz one by adding a near site which 
monitors the antineutrino flux from the two nuclear reactors at a distance 
of $\sim$410 m. The near site's overburden is 115 mwe. Double Chooz's far site 
is the original Chooz detector site whose baseline is 1,067 m and 
an overburden of 300 mwe. The total thermal power of the two
Double Chooz reactors is 8.7 GW. Figure~\ref{fig:dc-map-det} shows
the Double Chooz map and the detector design. The Double Chooz detector, like
all current-generation reactor antineutrino detectors, adapts a three-zone
design with the inner-most Gd-doped LS region as the target. Double Chooz
chooses PXE-based LS. The Gd doping is about 1~g/l. Its target mass is
10 t. Light from the target and the $\gamma$-catcher regions is monitored by
390 low-background 10-inch PMTs. 

Due to the delay in the civil construction, Double Chooz has so far only
collected far-site data. To constrain the reactor antineutrino flux
uncertainty, Double Chooz has used Bugey-4 measurement~\cite{Declais:1994ma} 
as the flux normalization. The Double Chooz analysis based on the 
neutron capture on Gd data gives
$\sin^22\theta_{13}=0.109\pm0.030~(\text{stat.})\pm0.025~(\text{sys.})$, 
which has considered
the prompt energy spectrum~\cite{Abe:2012tg}. Double Chooz has also carried out 
an independent $\theta_{13}$ analysis using the neutron capture on 
H data~\cite{Abe:2013sxa}. The H-capture measurement, 
$\sin^22\theta_{13}=0.097\pm0.034~(\text{stat.})\pm0.034~(\text{sys.})$, is
consistent with the Gd result. One advantage of Double Chooz
is its fewer number of reactors which can create a unique reactor off
data-taking condition. The direct background measurement during the 7.53
days of reactor-off period has enabled a background-independent
$\theta_{13}$ analysis~\cite{Abe:2014lus}. Combining the data of neutron 
captures on both Gd and H, Double Chooz measures
$\sin^22\theta_{13}=0.102\pm0.028~(\text{stat.})\pm0.033~(\text{sys.})$. 
The Double Chooz near detector is expected to start data taking in 
early 2014.

\section{Future Reactor Neutrino Experiments} % prospect 1 page % JUNO 3 pages
The current-generation reactor experiments will perform the ultimate 
measurement of $\bar{\nu}_e$ disappearance at a short baseline ($\sim$ 2 km).
Future reactor-based experiments will focus on the very
short baseline~(VSBL) and the medium baseline for different purposes. 
As examples, we pick one from each category, PROSPECT in the U.S. from VSBL
experiments and JUNO in China from medium-baseline experiments.
%In particular,
The PROSPECT experiment aims at resolving the
reactor anomaly~\cite{anom} at baselines $\sim$ 4-20 m. The JUNO experiment's 
major motivations includes the determination of the neutrino mass hierarchy 
and precision measurements of neutrino mixing parameters at baselines of
$\sim$53 km.

\subsection{PROSPECT}
PROSPECT (a Precision Reactor Oscillation and Spectrum Experiment at Very Short Baselines) 
is a multi-phased, multi-purposed, very short-baseline, research-reactor based,
neutrino experiment proposed in the U.S.~\cite{2013arXiv1309.7647A} The
collaboration is currently looking at three potential research reactor sites, 
the Advanced Test Reactor~(ATR) at the Idaho National Laboratory~(INL), 
the High Flux Isotope Reactor~(HFIR) at Oak Ridge National Laboratory~(ORNL), 
and the National Bureau of Standards Reactors~(NBSR) at National Institute of
Standards and Technology~(NIST). 
The research reactor sites generally allows baselines as short as a few meters,
which are the most interested region for the sterile neutrino search hinted by
the reactor anomaly~\cite{anom}. To enhance the sensitivity to an extra mass eigenstate
whose mass-squared splitting with the active states is at $\sim$1 eV$^2$,
PROSPECT collaboration adopts a segmented detector design to provide essential
resolutions in $L/E$. PROSPECT also has a unique phased approach. In its
first phase, a near detector within 10 m from the reactor core will be
installed and PROSPECT will cover $L/E$ in the range of 0.5-2.5 m/MeV. In its
second phase, PROSPECT will install a far detector with a baseline of 10-20 m,
which will extend the $L/E$ coverages to ~6 m/MeV. With these $L/E$ coverages,
PROSPECT will be able to exclude most of the parameter space allowed by the
reactor anomaly with high confidence levels. 

Besides its high quality data providing a definite test on the
reactor anomaly, PROSPECT data also has great potential in constraining reactor
antineutrino flux for other reactor neutrino experiments 
and for the nuclear non-proliferation industry. 
All three candidate research reactors of PROSPECT use highly-enriched
uranium~(HEU) whose antineutrinos are almost exclusively from $^{235}$U
fissions. The $^{235}$U antineutrino flux is the most precisely predicted
one based on the ILL beta spectrometer measurement~\cite{VonFeilitzsch:1982jw,Schreckenbach:1985ep}. 
Therefore, PROSPECT
will be able to provide an valuable benchmark to the reactor antineutrino flux
prediction and the reactor core simulation. Combined with existing flux
measurements at commercial reactors, PROSPECT data can also be used to test the
flux calculations other than $^{235}$U. Improved knowledge in the reactor
antineutrino flux prediction is going to be highly valuable to future
reactor based neutrino experiments. The high precision measurement of the
reactor antineutrino spectrum at a near-surface operation 
will also naturally benefit the development of  reactor safeguards. 

\subsection{Jiangmen Underground Neutrino Observatory} 
JUNO will be built in the Jiangmen City, Guangdong Province, 
China~\cite{Yifang:2012,Yifang:2013,Yifang:2013july}. The central 
piece of this experiment is a 20 kt liquid scintillator detector.
This detector will observe $\bar{\nu}_e$ from two reactor complexes: 
Taishan and Yangjiang. The Taishan reactor complex contains six reactor 
cores with a total thermal power of 17.4 GW. The Yangjiang reactor 
complex has two reactor cores with a total thermal power of 9.2 GW.
There are two additional reactor cores (9.2 GW) planned at the Yangjiang
site. The average baseline of JUNO is $\sim$52.5 km with a RMS (root 
mean square) of 0.25 km. The construction and the data 
taking are expected to start in 2015 and 2020, respectively.

\begin{figure}[htp]
\begin{centering}
\includegraphics[width=0.49\textwidth]{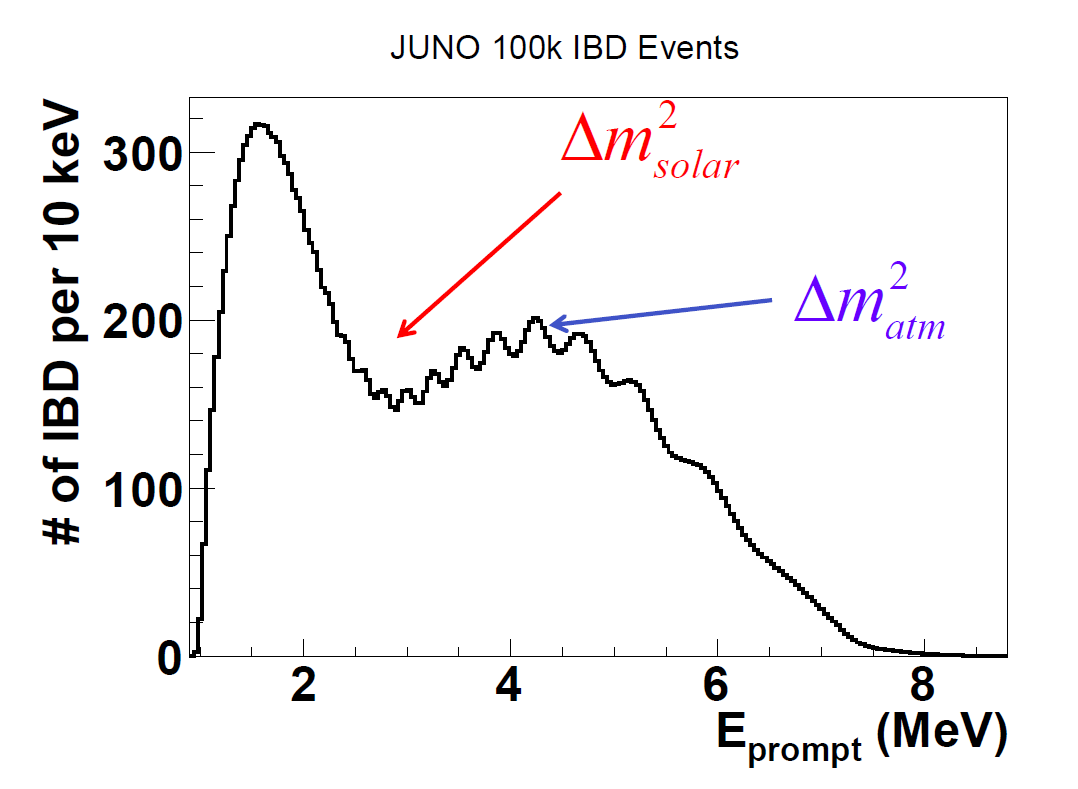}	
\includegraphics[width=0.49\textwidth]{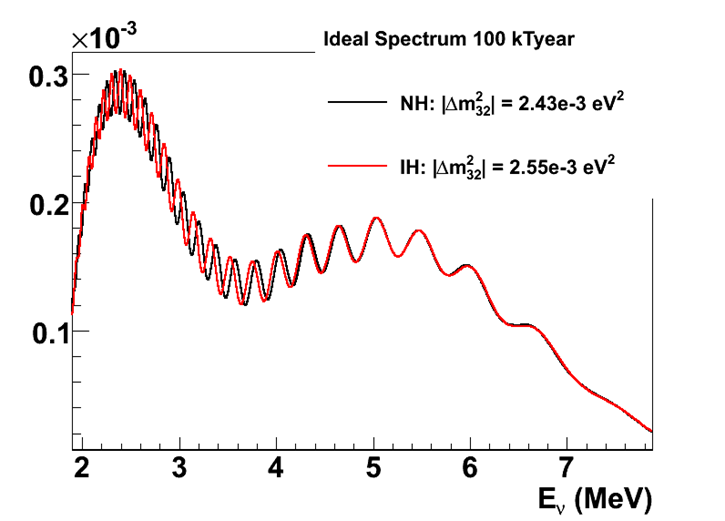}
\par\end{centering}
\caption{\label{fig:juno_spec} 
(Left) The expected nominal prompt energy spectrum of JUNO. 
A total of 100k IBD events, which corresponds to six years of data 
taking with a 20 kt detector and 36 GW$_{th}$ reactor power, is assumed. 
The big dip around 3 MeV corresponds to the solar oscillation 
($\Delta m^2_{21}$). The small wiggles from 2 to 8 MeV correspond 
to the atmospheric oscillation ($\Delta m^2_{ee}$). A $3\%/\sqrt{E~(MeV)}$ 
energy resolution is assumed. (Right) The ideal spectral distortion  
at JUNO (arbitrary scale in 
the vertical axis) for both normal and inverted hierarchies with a perfect energy 
resolution. Plots are taken from Ref.~\protect\cite{juno_run}.}
\end{figure}

Through the measurement of $\bar{\nu}_e$ disappearance at $\sim$53 km, 
JUNO's major physics goals are: i) the first experiment to simultaneously observe 
neutrino oscillations from both the atmospheric and the solar neutrino 
mass-squared splittings (see the left panel of Fig.~\ref{fig:juno_spec}),  
ii)  the first experiment to observe more than two oscillation 
cycles of the atmospheric mass-squared splitting (see the left panel 
of Fig.~\ref{fig:juno_spec}), 
iii) determination of the neutrino mass hierarchy, whether $\Delta m^2_{32}$
is larger or smaller than zero, through the measurement of the spectral distortion, and 
iv) precision measurements of $\sin^22\theta_{21}$, $\Delta m^2_{32}$, 
and $\Delta m^2_{21}$ to better than 1\%. We should note that 
the precision measurement of $\Delta m^2_{32}$ requires the 
knowledge of the neutrino mass hierarchy. Besides these, the 20 kt detector 
offers a rich physics program of the proton decay, geoneutrinos, 
supernova neutrinos, and many exotic neutrino physics topics.

\begin{figure}[htp]
\begin{centering}
\includegraphics[width=0.7\textwidth]{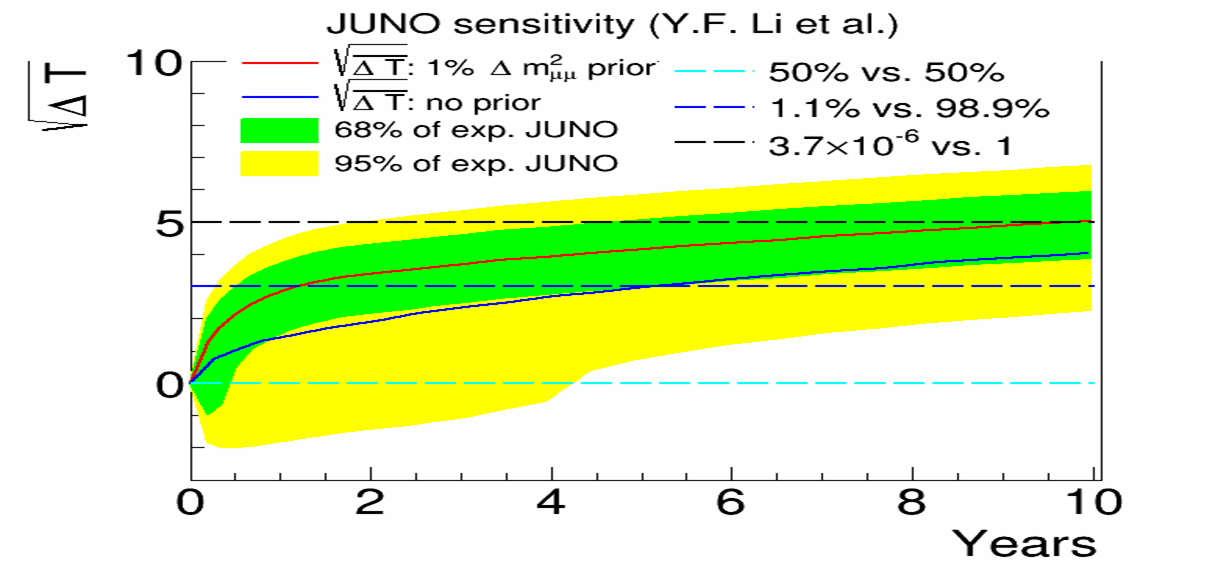}	
\par\end{centering}
\caption{\label{fig:juno_sen}  
JUNO's sensitivity evolution with respect to calendar years.~\protect\cite{Li:2013zyd}. 
A 20 kt detector at $\sim$ 53 km with a total of 36 GW$_{th}$ reactor power was 
assumed. The energy resolution was assumed to be $3\%/\sqrt{E~(MeV)}$. 
Plot is taken from Ref.~\protect\cite{juno_run}.}
\end{figure}

% MH ...
The neutrino mass hierarchy (MH) is likely to be the next determined 
fundamental parameter in the neutrino Standard Model. In combination
with searches for the neutrinoless double beta decay, the 
determination of MH will provide crucial information regarding the nature of neutrinos (whether they are Dirac or Majorana fermions). 
The non-zero $\theta_{13}$
established by the current-generation reactor experiments opened 
the path to determine the MH in a medium baseline ($\sim$55 km)
reactor experiment~\cite{Petcov:2001sy,Learned:2006wy,Zhan:2008id,Zhan:2009rs,Qian:2012xh,Ciuffoli:2012iz,Ge:2012wj,Li:2013zyd}. One simple 
way to understand the principle of MH determination is through 
the effective mass-squared splitting $\Delta m^2_{ee}$. At 
55 km baseline, the $\Delta m^2_{ee}$ measured at low energy 
($\sim$3 MeV) will be different from that measured at high 
energy ($\sim$6 MeV). For the normal MH, $\Delta m^2_{ee}$ at low energy
will be larger than that at high energy, and vice versa for 
IH.  The difference in the spectral distortion (with a perfect 
energy resolution) for NH and IH is shown in the right panel of 
Fig.~\ref{fig:juno_spec}. In order to reach this goal, JUNO requires
i) a better than $\sim$3\%$/ \sqrt{E(MeV)}$ energy resolution, 
ii) a high statistics IBD sample ($>$100k), iii) a $<$1\% absolute 
energy scale uncertainty~\cite{Qian:2012xh,Kettell:2013eos}. In 
addition, the site choice of JUNO was optimized taking into 
account the locations of reactor cores. Figure~\ref{fig:juno_sen} 
shows the expected sensitivity of JUNO~\cite{Li:2013zyd} with respect to the running 
time. The $\Delta T$ is a test statistics consisting likelihoods
of normal and inverted MH for data $x$. The green and yellow bands
represent the 68\% and 95\% expectations, respectively, taking into 
account the fluctuations in statistics and variations in systematics. 
The dotted lines correspond to the probability ratios of the normal 
vs. inverted MH in the Bayesian framework~\cite{Qian:2012zn}.

In addition to the determination of MH, JUNO will perform precision 
measurement of neutrino mixing, which is a powerful tool to test
the standard 3-flavor neutrino framework (or $\nu$SM). The precision
measurement of $\sin^22\theta_{12}$ will i) lay the foundation for 
a future sub-1\% direct unitarity test of the PMNS 
matrix~\cite{Antusch:2006vwa,Qian:2013ora},  
ii) constrain the allowed region of the effective neutrino mass to 
which the decay width of neutrinoless double beta decay is proportional,
and iii) test models of neutrino masses and mixing~\cite{King:2014nza}, 
such as 
$\theta_{12} = 35^\circ + \theta_{13}\cos\delta$, 
$\theta_{12} = 32^\circ + \theta_{13}\cos\delta$, and
$\theta_{12} = 45+\theta_{13}\cos\delta$. The precision measurement 
of $\Delta m^2_{ee}$ (or $\Delta m^2_{32}$) will
i) test an important sum rule  $\Delta m^2_{13} + \Delta m^2_{21} + \Delta m^2_{32} = 0$ and 
ii) reveal additional information regarding the neutrino mass hierarchy,
when combined with the precision $\Delta m^2_{\mu\mu}$ measurements 
from muon (anti)neutrino disappearance in accelerator experiments. 
As shown in Ref.~\cite{Caojun:2012}, the expected JUNO's precision of 
$\Delta m^2_{21}$, $\Delta m^2_{32}$, and $\sin^22\theta_{21}$ are 
0.3\%, 0.3\%, and 0.6\%, respectively. Such precision potential is 
further confirmed by 
studies made in Ref.~\cite{Kettell:2013eos}. 

% precision ..
% detector requirement ???
\begin{figure}[htp]
\begin{centering}
\includegraphics[width=0.9\textwidth]{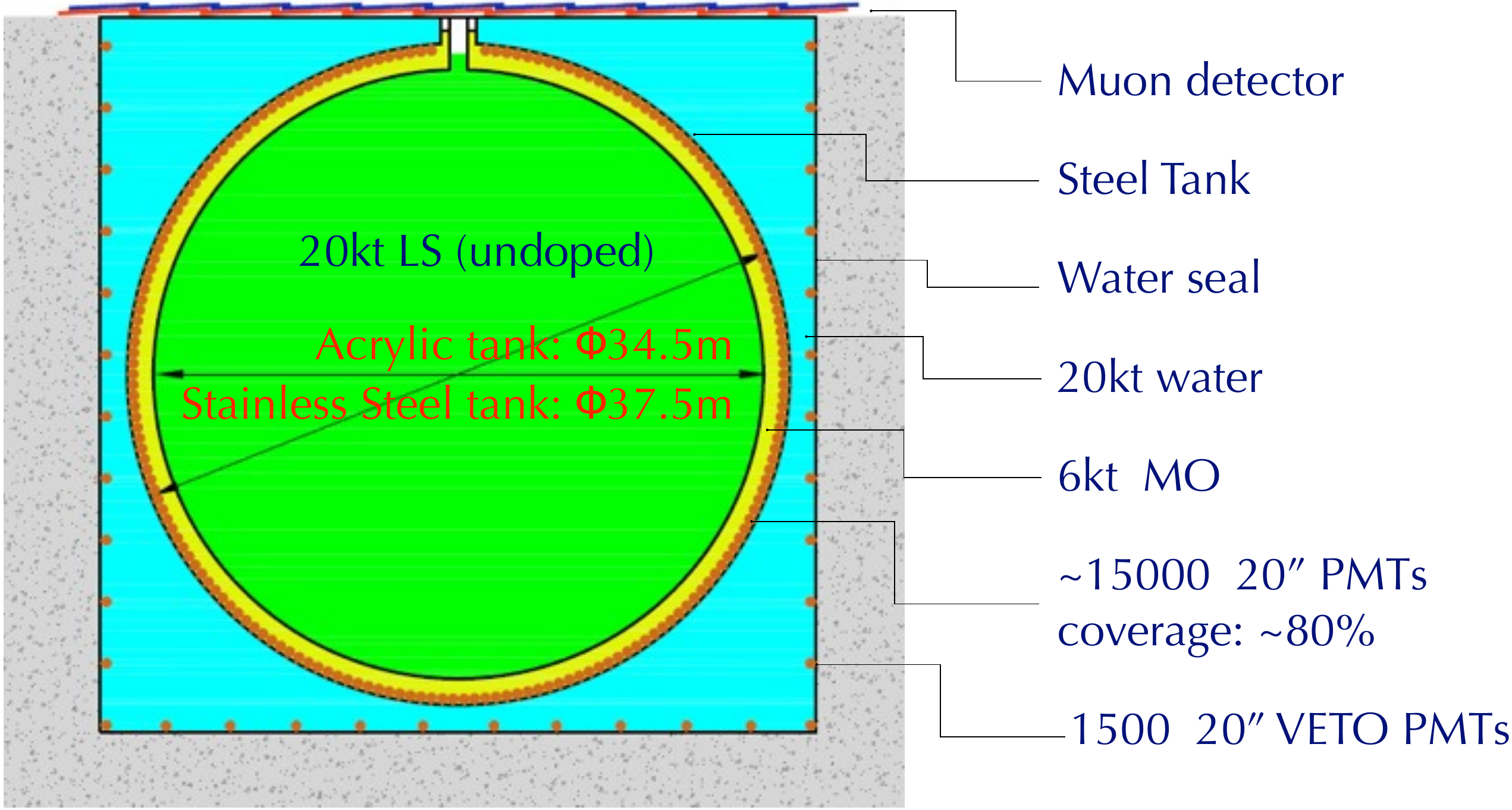}	
\par\end{centering}
\caption{\label{fig:juno_setup}One conceptual design of the JUNO detector. 
Unlike the current-generation short-baseline reactor antineutrino three-zone 
detectors, JUNO detector may adapt a two-zone design like the KamLAND one. 
And its target LS would be undoped due to the considerations of LS 
transparency and the unavoidable radioactive contamination in doping 
elements.}
\end{figure}

The central detector of JUNO will be a 20 kt underground 
liquid scintillator detector with a 1,850 m water equivalent overburden. 
Figure~\ref{fig:juno_setup} shows one conceptual design of JUNO's 20 kt LS 
detector~\cite{Yuekun:111,Yuekun:171}. A spherical LS target volume
is chosen i) to minimize the surface-to-volume ratio and PMT costs
and ii) to minimize position dependent corrections to the reconstructed
energy. The photo-cathode coverage is expected to reach $\sim$80\%.
Together with the high performance LS (high intrinsic photon yield  with 
$>$14,000 photons per MeV, the superior optical attenuation length of 
30 m or better) and the high quantum efficiency PMT, JUNO is aiming to 
achieve a better than $3\%/\sqrt{E~(MeV)}$ energy resolution that 
is essential for MH determination.

\section{Summary} % ~0.5 page
There were many discoveries in neutrino oscillation physics 
in the last decade. With the current-generation reactor experiments, we 
now know the value of $\theta_{13}$ 
(Daya Bay: $\sin^22\theta_{13} = 0.090^{+0.008}_{-0.009}$). The large
value of $\theta_{13}$ opens doors to access remaining unknowns 
in the $\nu$SM: the neutrino mass hierarchy and the leptoic CP 
phase $\delta_{CP}$. In particular, the next generation (medium-baseline) 
reactor experiments aims to resolve the neutrino mass hierarchy. 
As we enter the precision era of neutrino physics, the current and
future reactor experiments will bring us more exciting findings.

\section*{Acknowledgments}
We would like to thank Zhizhong Xing, Dmitry Naumov, Karsten Heeger for
helpful inputs. This work was supported in part by the National Science 
Foundation and the Department of Energy under contracts DE-AC02-98CH10886.
\bibliographystyle{unsrt}
\bibliography{dyb_review_mpla}{}

\begin{thebibliography}{10}

\bibitem{nu_dis}
C.L. Cowan, F.~Reines, F.B. Harrison, H.W. Kruse, and A.D. McGuire.
\newblock {Detection of the free neutrino: A Confirmation}.
\newblock {\em Science}, 124:103--104, 1956.

\bibitem{KamLAND_spec}
T.~Araki et~al.
\newblock {Measurement of neutrino oscillation with KamLAND: Evidence of
  spectral distortion}.
\newblock {\em Phys.Rev.Lett.}, 94:081801, 2005.

\bibitem{SNO}
Q.R. Ahmad et~al.
\newblock {Direct evidence for neutrino flavor transformation from neutral
  current interactions in the Sudbury Neutrino Observatory}.
\newblock {\em Phys.Rev.Lett.}, 89:011301, 2002.

\bibitem{dayabay}
F.P. An et~al.
\newblock {Observation of electron-antineutrino disappearance at Daya Bay}.
\newblock {\em Phys.Rev.Lett.}, 108:171803, 2012.

\bibitem{ponte1}
B.~Pontecorvo.
\newblock {Mesonium and anti-mesonium}.
\newblock {\em Sov.Phys.JETP}, 6:429, 1957.

\bibitem{ponte2}
B.~Pontecorvo.
\newblock {Neutrino Experiments and the Problem of Conservation of Leptonic
  Charge}.
\newblock {\em Sov.Phys.JETP}, 26:984--988, 1968.

\bibitem{Maki}
Ziro Maki, Masami Nakagawa, and Shoichi Sakata.
\newblock {Remarks on the unified model of elementary particles}.
\newblock {\em Prog.Theor.Phys.}, 28:870--880, 1962.

\bibitem{PDG}
J.~Beringer et~al.
\newblock {Review of Particle Physics (RPP)}.
\newblock {\em Phys.Rev.}, D86:010001, 2012.

\bibitem{Nuno}
H.~Minakata, H.~Nunokawa, Stephen~J. Parke, and R.~Zukanovich~Funchal.
\newblock {Determining neutrino mass hierarchy by precision measurements in
  electron and muon neutrino disappearance experiments}.
\newblock {\em Phys.Rev.}, D74:053008, 2006.

\bibitem{Chooz1}
M.~Apollonio et~al.
\newblock {Limits on neutrino oscillations from the CHOOZ experiment}.
\newblock {\em Phys.Lett.}, B466:415--430, 1999.

\bibitem{Chooz2}
M.~Apollonio et~al.
\newblock {Search for neutrino oscillations on a long baseline at the CHOOZ
  nuclear power station}.
\newblock {\em Eur.Phys.J.}, C27:331--374, 2003.

\bibitem{PaloVerde}
F.~Boehm, J.~Busenitz, B.~Cook, G.~Gratta, H.~Henrikson, et~al.
\newblock {Final results from the Palo Verde neutrino oscillation experiment}.
\newblock {\em Phys.Rev.}, D64:112001, 2001.

\bibitem{Fogli:2011qn}
G.L. Fogli, E.~Lisi, A.~Marrone, A.~Palazzo, and A.M. Rotunno.
\newblock {Evidence of $\theta_{13}>0$ from global neutrino data analysis}.
\newblock {\em Phys.Rev.}, D84:053007, 2011.

\bibitem{minos}
P.~Adamson et~al.
\newblock {Improved search for muon-neutrino to electron-neutrino oscillations
  in MINOS}.
\newblock {\em Phys.Rev.Lett.}, 107:181802, 2011.

\bibitem{t2k}
K.~Abe et~al.
\newblock {Indication of Electron Neutrino Appearance from an
  Accelerator-produced Off-axis Muon Neutrino Beam}.
\newblock {\em Phys.Rev.Lett.}, 107:041801, 2011.

\bibitem{dc}
Y.~Abe et~al.
\newblock {Indication for the disappearance of reactor electron antineutrinos
  in the Double Chooz experiment}.
\newblock {\em Phys.Rev.Lett.}, 108:131801, 2012.

\bibitem{RENO}
J.K. Ahn et~al.
\newblock {Observation of Reactor Electron Antineutrino Disappearance in the
  RENO Experiment}.
\newblock {\em Phys.Rev.Lett.}, 108:191802, 2012.

\bibitem{Evans:2013pka}
Justin Evans.
\newblock {The MINOS experiment: results and prospects}.
\newblock {\em Adv.High Energy Phys.}, 2013:182537, 2013.

\bibitem{Abe:2013hdq}
K.~Abe et~al.
\newblock {Observation of Electron Neutrino Appearance in a Muon Neutrino
  Beam}.
\newblock 2013.

\bibitem{Abe:2013sxa}
Y.~Abe et~al.
\newblock {First Measurement of $\theta_{13}$ from Delayed Neutron Capture on
  Hydrogen in the Double Chooz Experiment}.
\newblock {\em Phys.Lett.}, B723:66--70, 2013.

\bibitem{2013arXiv1312.4111S}
S.-H. {Seo}.
\newblock {New Results from RENO}.
\newblock {\em ArXiv e-prints}, December 2013.

\bibitem{An:2013zwz}
F.P. An et~al.
\newblock {Spectral measurement of electron antineutrino oscillation amplitude
  and frequency at Daya Bay}.
\newblock 2013.

\bibitem{ratio}
L.A. Mikaelyan and V.V. Sinev.
\newblock {Neutrino oscillations at reactors: What next?}
\newblock {\em Phys.Atom.Nucl.}, 63:1002--1006, 2000.

\bibitem{anom}
G.~Mention, M.~Fechner, Th. Lasserre, Th.A. Mueller, D.~Lhuillier, et~al.
\newblock {The Reactor Antineutrino Anomaly}.
\newblock {\em Phys.Rev.}, D83:073006, 2011.

\bibitem{anom1}
C.~Zhang, X.~Qian, and P.~Vogel.
\newblock {Reactor Antineutrino Anomaly with known $\theta_{13}$}.
\newblock {\em Phys.Rev.}, D87(7):073018, 2013.

\bibitem{anom2}
A.C. Hayes, J.L. Friar, G.T. Garvey, and Guy Jonkmans.
\newblock {Reanalysis of the Reactor Neutrino Anomaly}.
\newblock 2013.

\bibitem{Band:2013osa}
H.R. Band, R.~Carr, X.C. Chen, X.H. Chen, J.J. Cherwinka, et~al.
\newblock {Assembly and Installation of the Daya Bay Antineutrino Detectors}.
\newblock {\em JINST}, 8:T11006, 2013.

\bibitem{Liu:2013ava}
J.~Liu, B.~Cai, R.~Carr, D.A. Dwyer, W.Q. Gu, et~al.
\newblock {Automated Calibration System for a High-Precision Measurement of
  Neutrino Mixing Angle $\theta_{13}$ with the Daya Bay Antineutrino
  Detectors}.
\newblock 2013.

\bibitem{DayaBay:2012aa}
F.P. An et~al.
\newblock {A side-by-side comparison of Daya Bay antineutrino detectors}.
\newblock {\em Nucl.Instrum.Meth.}, A685:78--97, 2012.

\bibitem{minos_update}
P.~Adamson et~al.
\newblock {Measurement of Neutrino and Antineutrino Oscillations Using Beam and
  Atmospheric Data in MINOS}.
\newblock {\em Phys.Rev.Lett.}, 110:251801, 2013.

\bibitem{VonFeilitzsch:1982jw}
F.~Von~Feilitzsch, A.A. Hahn, and K.~Schreckenbach.
\newblock {EXPERIMENTAL BETA SPECTRA FROM PU-239 AND U-235 THERMAL NEUTRON
  FISSION PRODUCTS AND THEIR CORRELATED ANTI-NEUTRINOS SPECTRA}.
\newblock {\em Phys.Lett.}, B118:162--166, 1982.

\bibitem{Schreckenbach:1985ep}
K.~Schreckenbach, G.~Colvin, W.~Gelletly, and F.~Von~Feilitzsch.
\newblock {DETERMINATION OF THE ANTI-NEUTRINO SPECTRUM FROM U-235 THERMAL
  NEUTRON FISSION PRODUCTS UP TO 9.5-MEV}.
\newblock {\em Phys.Lett.}, B160:325--330, 1985.

\bibitem{Hahn:1989zr}
A.A. Hahn, K.~Schreckenbach, G.~Colvin, B.~Krusche, W.~Gelletly, et~al.
\newblock {Anti-neutrino Spectra From $^{241}$Pu and $^{239}$Pu Thermal Neutron
  Fission Products}.
\newblock {\em Phys.Lett.}, B218:365--368, 1989.

\bibitem{Huber:2011wv}
Patrick Huber.
\newblock {On the determination of anti-neutrino spectra from nuclear
  reactors}.
\newblock {\em Phys.Rev.}, C84:024617, 2011.

\bibitem{Vogel:1980bk}
P.~Vogel, G.K. Schenter, F.M. Mann, and R.E. Schenter.
\newblock {Reactor Anti-neutrino Spectra and Their Application to Anti-neutrino
  Induced Reactions. 2.}
\newblock {\em Phys.Rev.}, C24:1543--1553, 1981.

\bibitem{Mueller:2011nm}
Th.A. Mueller, D.~Lhuillier, M.~Fallot, A.~Letourneau, S.~Cormon, et~al.
\newblock {Improved Predictions of Reactor Antineutrino Spectra}.
\newblock {\em Phys.Rev.}, C83:054615, 2011.

\bibitem{dyb_run}
Daya~Bay Collaboration.
\newblock U.S. P5 contribution,
  \url{http://www.usparticlephysics.org/sites/default/files/webform/p5/DayaBay%
_RunPlan_0.pdf} (2013).

\bibitem{Adams:2013qkq}
C.~Adams et~al.
\newblock {Scientific Opportunities with the Long-Baseline Neutrino
  Experiment}.
\newblock 2013.

\bibitem{Qian:2013ora}
X.~Qian, C.~Zhang, M.~Diwan, and P.~Vogel.
\newblock {Unitarity Tests of the Neutrino Mixing Matrix}.
\newblock 2013.

\bibitem{Declais:1994ma}
Y.~Declais, H.~de~Kerret, B.~Lefievre, M.~Obolensky, A.~Etenko, et~al.
\newblock {Study of reactor anti-neutrino interaction with proton at Bugey
  nuclear power plant}.
\newblock {\em Phys.Lett.}, B338:383--389, 1994.

\bibitem{Abe:2012tg}
Y.~Abe et~al.
\newblock {Reactor electron antineutrino disappearance in the Double Chooz
  experiment}.
\newblock {\em Phys.Rev.}, D86:052008, 2012.

\bibitem{Abe:2014lus}
Y.~Abe et~al.
\newblock {Background-independent measurement of $\theta_{13}$ in Double
  Chooz}.
\newblock 2014.

\bibitem{2013arXiv1309.7647A}
J.~{Ashenfelter}, A.~B. {Balantekin}, H.~{Band}, A.~{Bernstein}, E.~{Blucher},
  N.~S. {Bowden}, C.~{Bryan}, J.~C. {Cherwinka}, T.~{Classen}, D.~{Dean}, M.~J.
  {Dolinski}, Y.~{Efremenko}, A.~{Galindo-Uribarri}, A.~{Glenn}, M.~{Green},
  S.~{Hans}, K.~M. {Heeger}, R.~{Henning}, L.~{Hu}, P.~{Huber}, R.~{Johnson},
  C.~{Lane}, T.~J. {Langford}, J.~G. {Learned}, B.~R. {Littlejohn},
  J.~{Maricic}, B.~{McKeown}, S.~{Morrell}, H.~P. {Mumm}, J.~S. {Nico},
  E.~{Romero}, S.~J. {Thompson}, W.~{Wang}, B.~{White}, R.~E. {Williams},
  T.~{Wise}, M.~{Yeh}, and N.~{Zaitseva}.
\newblock {PROSPECT - A Precision Reactor Oscillation and Spectrum Experiment
  at Very Short Baselines}.
\newblock {\em ArXiv e-prints}, September 2013.

\bibitem{Yifang:2012}
Yifang Wang.
\newblock {Daya Bay II}: the next generation reactor neutrino experiment.
\newblock In {\em NuFACT 2012}, Jul 2012.

\bibitem{Yifang:2013}
Yifang Wang.
\newblock {Daya Bay II}: current status and future plan.
\newblock In {\em Daya Bay II First Meeting}, Jan 2013.

\bibitem{Yifang:2013july}
Yifang Wang.
\newblock {JUNO}: current status and future plan.
\newblock In {\em JUNO Meeting at IHEP, Beijing, China}, Jul 2013.

\bibitem{juno_run}
MBRO Collaboration.
\newblock U.S. P5 contribution,
  \url{http://www.usparticlephysics.org/sites/default/files/webform/p5/US_JUNO%
_P5.pdf} (2014).

\bibitem{Li:2013zyd}
Yu-Feng Li, Jun Cao, Yifang Wang, and Liang Zhan.
\newblock {Unambiguous Determination of the Neutrino Mass Hierarchy Using
  Reactor Neutrinos}.
\newblock {\em Phys.Rev.}, D88:013008, 2013.

\bibitem{Petcov:2001sy}
S.T. Petcov and M.~Piai.
\newblock {The LMA MSW solution of the solar neutrino problem, inverted
  neutrino mass hierarchy and reactor neutrino experiments}.
\newblock {\em Phys.Lett.}, B533:94--106, 2002.

\bibitem{Learned:2006wy}
John Learned, Stephen~T. Dye, Sandip Pakvasa, and Robert~C. Svoboda.
\newblock {Determination of neutrino mass hierarchy and $\theta_{13}$ with a
  remote detector of reactor antineutrinos}.
\newblock {\em Phys.Rev.}, D78:071302, 2008.

\bibitem{Zhan:2008id}
Liang Zhan, Yifang Wang, Jun Cao, and Liangjian Wen.
\newblock {Determination of the Neutrino Mass Hierarchy at an Intermediate
  Baseline}.
\newblock {\em Phys.Rev.}, D78:111103, 2008.

\bibitem{Zhan:2009rs}
Liang Zhan, Yifang Wang, Jun Cao, and Liangjian Wen.
\newblock {Experimental Requirements to Determine the Neutrino Mass Hierarchy
  Using Reactor Neutrinos}.
\newblock {\em Phys.Rev.}, D79:073007, 2009.

\bibitem{Qian:2012xh}
X.~Qian, D.A. Dwyer, R.D. McKeown, P.~Vogel, W.~Wang, et~al.
\newblock {Mass Hierarchy Resolution in Reactor Anti-neutrino Experiments:
  Parameter Degeneracies and Detector Energy Response}.
\newblock {\em PRD, 87,}, 033005, 2013.

\bibitem{Ciuffoli:2012iz}
Emilio Ciuffoli, Jarah Evslin, and Xinmin Zhang.
\newblock {The Neutrino Mass Hierarchy at Reactor Experiments now that theta13
  is Large}.
\newblock {\em JHEP}, 1303:016, 2013.

\bibitem{Ge:2012wj}
Shao-Feng Ge, Kaoru Hagiwara, Naotoshi Okamura, and Yoshitaro Takaesu.
\newblock {Determination of mass hierarchy with medium baseline reactor
  neutrino experiments}.
\newblock {\em JHEP}, 1305:131, 2013.

\bibitem{Kettell:2013eos}
A.~B. Balantekin et~al.
\newblock {Neutrino mass hierarchy determination and other physics potential of
  medium-baseline reactor neutrino oscillation experiments}.
\newblock 2013.

\bibitem{Qian:2012zn}
X.~Qian, A.~Tan, W.~Wang, J.J. Ling, R.D. McKeown, et~al.
\newblock {Statistical Evaluation of Experimental Determinations of Neutrino
  Mass Hierarchy}.
\newblock {\em Phys.Rev.}, D86:113011, 2012.

\bibitem{Antusch:2006vwa}
S.~Antusch, C.~Biggio, E.~Fernandez-Martinez, M.B. Gavela, and J.~Lopez-Pavon.
\newblock {Unitarity of the Leptonic Mixing Matrix}.
\newblock {\em JHEP}, 0610:084, 2006.

\bibitem{King:2014nza}
Stephen~F. King, Alexander Merle, Stefano Morisi, Yusuke Shimizu, and Morimitsu
  Tanimoto.
\newblock {Neutrino Mass and Mixing: from Theory to Experiment}.
\newblock 2014.

\bibitem{Caojun:2012}
Jun Cao.
\newblock Daya bay-ii: A 60 km-baseline reactor experiment and beyond.
\newblock In {\em International Symposium on Neutrino Physics and Beyond},
  2012.

\bibitem{Yuekun:111}
Yue-Kun Heng.
\newblock The central detector design and acrylic option.
\newblock Jul 2013.

\bibitem{Yuekun:171}
Yue-Kun Heng.
\newblock The progress of the central detector.
\newblock Oct 2013.

\end{thebibliography}

%\begin{thebibliography}{0}
%\end{thebibliography}

\end{document}